\documentclass[pre,aps,twocolumn,floatfix,showpacs]{revtex4}
\usepackage{dcolumn}
\usepackage{graphicx}
\usepackage{graphicx,amssymb,amsmath}
\usepackage{natbib}
\usepackage{bm}

\begin{document}
\title{The nonlinear elasticity of an
$\alpha$-helical polypeptide: Monte Carlo studies}
\author{Buddhapriya Chakrabarti}
\email{buddho@physics.umass.edu}
\author{Alex J. Levine}
\email{levine@physics.umass.edu} \affiliation{Department of
Physics, University of Massachusetts.\\ Amherst, MA 01003}
\date{\today}

\begin{abstract}
We perform Monte Carlo simulations to study the elastic properties
of the helix-coil worm-like chain model of alpha-helical
polypeptides. In this model the secondary structure enters as a
scalar (Ising like) variable that controls the local chain bending
modulus. We compute numerically the bending and stretching
compliances of this molecule as well as the nonlinear interaction
between stretching and torque over a wide range of model
parameters. The numerical results agree well with previous
mean-field and perturbative calculations where they are expected
to do so. The Monte Carlo simulations allow us to examine the
response of the chain to large forces and torques where the
perturbative approaches fail. In addition we extend our mean-field
analysis by studying the fluctuation dominated regime at the
force-induced denaturation transition.
\end{abstract}
 \maketitle

\section{Introduction}
The mechanical properties of semiflexible biopolymers such as
F-actin are well described by the worm-like chain
Hamiltonian\cite{Kratky:49,Fisher:63}, which treats the filament
as a one dimensional elastic continuum without internal structure.
Experiments based on single molecule
manipulations\cite{Smith:92,Perkins:94,Perkins:95,Cluzel:96,Strick:96,Perkins:97,Bustamante:03,Strick:03}
have demonstrated the essential validity of this coarse-grained
approach to the investigation of biopolymer statistics and
mechanical properties. Understanding both scattering function of
such semiflexible polymers in dilute solution\cite{Benoit:53} and
the force extension relations\cite{Marko:95} of single
semiflexible polymers are among the principal successes of this
coarse-grained description of these polymers. A simplification of
this nature, although valid in numerous contexts, must breakdown
under forces that are large enough to modify the local structure
of the polymer. For example the worm-like chain Hamiltonian
ignores the double helical structure of DNA and the local
secondary structure of polypeptides. This local molecular
structure controls the bending modulus of the polymer and can be
disrupted under applied stress. This limitation of the worm-like
chain (WLC) model is becoming increasingly apparent as researchers
probe the mechanical properties of biopolymers under larger
applied forces where such a coupling between local molecular
structure and chain elasticity plays an important role
\cite{Rief:90}.  Such large forces are not only experimentally
accessible but are also biologically relevant in such processes as
those associated with DNA looping\cite{Coutier:04,Nelson:04} and
in protein conformational change. An example of the latter process
can be found in the conformational change of the protein
Calmodulin that involves the buckling of a single, solvent-exposed
alpha-helical domain upon the binding of Ca$^{2+}$
ions\cite{calmodulin,Wriggers:98}. A polymer model of the
alpha-helical domain that incorporates this nonlinear elastic
response of the chain is therefore required to explore
conformation change. We expect that such a model (as presented
here) will be broadly applicable to the investigation of protein
allostery.

Recently we\cite{Levine:04,Chakrabarti:04} and
others\cite{Storm:03,Nelson:04} proposed a minimal extension of
worm-like chain called the helix/coil worm-like chain (HCWLC) to
describe biopolymers with internal structure. This model couples
the conformational degrees of freedom of the polymer backbone to
localized structural transitions of the constituent monomers by
postulating that the local bending modulus of the worm-like chain
depends on the local degree of internal structure (the helix/coil
variable). For example, the bending modulus of an alpha-helical
polypeptide depends on the local presence of secondary structure
and the consequent hydrogen bonding that effectively stiffens the
chain. The fundamental result of this coupling is to make both the
torque and force response of the polymer highly nonlinear due to
localized denaturation events (loss of local secondary structure)
under mechanical stress. These denatured regions introduce more
compliant elements into the chain leading to abrupt changes in the
effective moduli of the polymer at a critical applied stress.

It is important to note that the secondary structure variables are
locally coupled along the chain since the alpha-helical structure
is stabilized by hydrogen bonding between adjacent turns of the
helix. These nearest neighbor interactions between the secondary
structure variables make the denaturation transition cooperative.
We will parameterize this cooperativity by an energy scale in what
follows referring to it as the chain cooperativity parameter. The
worm-like chain Hamiltonian and our extension of it, the HCWLC are
discussed in more detail in section~\ref{model}.

The numerical work reported here complements our previous analytic
calculations of the extension and bending compliances of
alpha-helical  polypeptides as modelled within the framework of
the HCWLC. For the case of externally applied torques with no
externally applied tensile stress, these analytic calculations
have been carried out exactly. Both the force extension relations
of the molecule and the nonlinear coupling of the extensional
compliance to the state of torque, however, cannot be solved in
closed form. A similar issue arises in the study of the
extensional compliance of the worm-like chain\cite{Marko:95}.
Previously we examined the response of the molecule to tensile
stress as well as to a combination of torque and tensile stress by
two routes: (i) perturbation theory applicable to the low force
regime, and (ii) mean field theory that replaces fluctuating
internal variables along the chain by their self-consistent, mean
value. This latter approximation is of particular utility for the
case of high chain cooperativity that suppresses strong
fluctuations of the secondary structure variables.  Here we
provide a Ginzburg criterion\cite{Chaikin:98} to discuss the
limits of validity of this approximation and use Monte Carlo
simulations to evaluate the response of the molecule in those
regimes where the previous mean-field or perturbative approaches
do not apply. We also use the Monte Carlo method to confirm our
previous calculations where such analytical calculations are
applicable.

This work thus explores a new range in the parameter space of the
model, {\em i.e.} the regime of strong secondary structure
fluctuations. Although estimates\cite{Chakrabarti:04} show that at
least some alpha-helical polypeptides under physiological
conditions have high chain cooperativity and thus are
well-approximated by the mean field description, exploring the
elastic properties of a polypeptide with low chain cooperativity
may well reflect the effects of poor solvent quality. In addition
we will show that the model predicts that there is always a
fluctuation--dominated regime near the point at which the
alpha-helix denatures under tension.

To briefly review the phenomenology of the model, we find a highly
nonlinear elastic response of the alpha-helix when subjected to
bending torques and stretching forces as a result of the
interaction between the local secondary structure and the
conformational fluctuations. In particular we reported a
mechanical instability of the chain under bending akin to a
``buckling instability'' in which the segment of alpha-helical
polypeptide locally melts to form a denatured (random coil)
segment beyond a critical angle. At this point the torque required
to hold the chain at that particular angle drops precipitously.

Upon application of a tensile stress to an alpha-helical polymer
we find three regimes of response. At the lowest stresses, the
external force pulls out the small equilibrium undulations of the
stiff alpha-helical polymer in a manner identical to the WLC.
There  are however, significant differences between the force
extension  behaviors of the WLC and the HCWLC. In the WLC model
the extensional  compliance of the chain vanishes in the limit of
large forces  because each segment of the chain is inextensible.
The applied tensile  force simply depletes the equilibrium
population of transverse  undulations along the chain. As these
undulations vanish no additional arc length can be recovered by
additional force; the measured extension of the chain plateaus at
its total contour length. This is not the case for the HCWLC. For
short enough chains (to be quantified below) the standard
Marko-Siggia plateau in which additional force does not produce
extension is replaced by what we term a ``pseudoplateau''
characterized by the fluctuations of the polymer into the random
coil, or denatured state. Since the random coil sections of the
chain are longer than the same segments in their native,
alpha-helical state, the biasing of the fluctuations into the
random coil leads to additional contour length of the polymer.
Finally in the third regime the chain is completely denatured by
the applied force and reaches a true Marko-Siggia plateau upon
further increasing the force.

The remainder of the paper is organized as follows. In section
\ref{model} we introduce the alpha-helix Hamiltonian based on a
combination of the worm-like chain and the helix/coil model. The
details of the Monte Carlo simulation is described in
\ref{montecarlo}. The simulation results are analyzed in the
section \ref{results}. We first study the radius of gyration of
the alpha helix as a function of solvent quality (to be defined
therein) in section \ref{radius-gyration}. We then examine the
response of the chain to bending torques in section \ref{bending}.
In section \ref{stretching} we study the extensional compliance of
the alpha-helix before summarizing and discussing the results in
section~\ref{conclusions}.

\section{The Helix-Coil Worm-like Chain model}
\label{model}

The worm-like chain (WLC)\cite{Kratky:49,Fisher:63} is the
simplest coarse-grained model for semiflexible polymers. This
model describes the single-chain polymer statistics in terms of a
Hamiltonian that associates an energy cost with chain curvature by
introducing a bending modulus $\kappa$. In terms of a discretized
chain model described by the set of monomeric tangent vectors
$\hat{t}_i$, $i = 0, \ldots, N-1$ with $N$ the degree of
polymerization, the WLC Hamiltonian may be written as
\begin{equation}
\label{WLC-Hamiltonian} H_{\rm WLC} = \kappa \sum_{i=0}^{N-1}
\left[ 1 - \left(\hat{t}_i \cdot \hat{t}_{i+1} \right) \right].
\end{equation}
The bending modulus $\kappa$ determines the thermal persistence
length of the chain, the typical distance along the chain over
which the tangent vectors decorrelate.

The response of a single chain to extensional forces has been used
to understand the deformational properties of biopolymers and
their aggregates\cite{Marko:95, MacKintosh:95,Lamura:01}. To
account for the internal degrees of freedom along the chain a new
set of variables is needed. Workers have previously employed the
helix/coil (HC) model\cite{Poland:70} in order to introduce such
internal state variables along the arc length of the chain. This
model has been used to study a class of protein conformational
transitions\cite{Birshtein:66,Bloomfield:99} in solution and under
tension\cite{Tamashiro:01}.

\begin{figure}
\includegraphics[width=7cm]{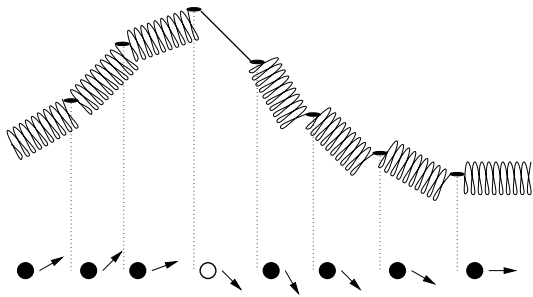}
\caption{\label{finalschem} Schematic figure of an alpha-helical
polypeptide and its representation in terms of the Ising-like
secondary structure variables (open circles for random coil
segments and filled ones for alpha-helical ones) and the tangent
vectors to the segments of the chain (denoted by arrows).}
\end{figure}

The HC model Hamiltonian, which is used to study these structural
transitions can be reduced to its simplest form by assuming that
the local structure of the chain can be described by a set of
two-state variables $s_i = \pm 1$, $ i = 0, \ldots, N$. For the
alpha-helical chains of current interest we regard these two
states as the local conformation of the segment in its native,
alpha-helical state ($s = +1$) and in a disordered, random coil
state ($s = -1$).

It is important to note that the elementary units of the chain as
described by the HCWLC model are not the amino acid monomers but
rather turns of the alpha helix since it is necessary to
unambiguously ascribe the presence or absence of secondary
structure to each elementary unit of the model. In practice we
expect this level of coarse graining to mean that each segment of
the chain (or elementary unit) is composed of $\sim 3$ monomers.

The coupling of the secondary structure variables to the WLC tangent
vectors is effected by introducing a bending stiffness in the WLC
Hamiltonian that depends on the local degree of secondary
structure. We choose
\begin{equation}
\label{kappa-def} \kappa(s) = \left\{
                \begin{array}{ll}
                        \kappa_> & \mbox{if $s = + 1$} \\
                        \kappa_< & \mbox{if $s=-1$}
                \end{array}
            \right. .
\end{equation}
Due to the hydrogen bonding between turns of the alpha-helix, it
is reasonable to expect that $\kappa_>$ the bending modulus in the
native state is significantly larger than $\kappa_<$ the bending
modulus of the chain in the non-native, disordered state. Simple
estimates of this difference in bending moduli were computed in
\cite{Chakrabarti:04}.

The Hamiltonian of the helix-coil worm-like chain may be thus
written as
\begin{eqnarray}
 H = \epsilon_{\rm w}/2 \sum^{N-1}_{i=0} (1 - s_i s_{i+1}) - h/2
\sum^{N}_{i=0} (s_i - 1)+ \nonumber \\
+ \sum^{N-1}_{i=0} \kappa(s_{i}) \left[ 1 - \left(\hat{t}_i \cdot
\hat{t}_{i+1} \right) \right], \label{HCWLC-hamiltonian}
\end{eqnarray}
where $\epsilon_{\rm w}$ is the free energy cost of a domain wall
in the secondary structure sequence and is the negative natural
logarithm of the chain cooperativity parameter; $h$ represents the
free energy cost per monomer to be in the non-native (\emph{i.e.}
random-coil) state, while $\kappa_>$ and $\kappa_<$ are the
bending moduli of the chain in the helix and coil phases as
mentioned above.  A pictorial representation of the system is
shown in figure \ref{finalschem}.

The full Hamiltonian given by Eq.~\ref{HCWLC-hamiltonian} has four
constants with dimensions of energy: $\kappa_>, \kappa_<, h,
\epsilon_{\rm w}$ that can be fit from experiment. We have
disregarded the twist degree of freedom of the molecule. Such
twist degrees of freedom and the coupling of twisting and
stretching modes of these chiral molecules have been explored
particularly with regard to the mechanical properties of
DNA\cite{Kamien:97,Ohern:98}. Also in the present work we evaluate
the model in two dimensions. Full three dimensional variants of
the calculation that incorporate torsional modes as well as the
twist-stretch coupling are currently under
investigation\cite{Chakrabarti:04a}.

\section{Monte Carlo simulations}
\label{montecarlo}

We performed Monte Carlo simulations of the helix coil worm--like
chain model using a standard Metropolis algorithm. Our system
consists of a linear chain of $N$ lattice points with two
variables specified at each lattice position, a scalar variable
$s_i = \pm 1$ recording the presence or absence of secondary
structure and an angular variable $\theta_i$ specifying the angle
the tangent vector joining two sites of the chain make with the
mean direction. The number of tangent vectors in the simulation is
one less than the total number of spins, {\em i.e.} it is $N-1$.

The energy of the chain is specified by the Hamiltonian given by
Eq.~\ref{HCWLC-hamiltonian}. We choose for our initial
configuration a state having equal numbers of helix and coil
segments and with randomized tangent vectors. There are two
classes of Monte Carlo moves: spin flips representing attempts to
change the local secondary structure and tangent vector moves
allowing the chain to explore all conformations. To update the
configuration of the chain, we flip a spin at random $s_i
\rightarrow - s_i$ and calculate the energy difference between the
final ($\nu$) and the initial configurations ($\mu$). The
acceptance probability of this move is given by
\begin{equation}
\label{metropolis} A(\mu \rightarrow \nu) = \left\{
\begin{array}{ll}
                        \exp[- \beta (E_{\nu} - E_{\mu}) ], & \mbox{
$E_{\nu}-E_{\mu}>0$} \\
                        1  & \mbox{otherwise}
                   \end{array}
            \right. ,
\end{equation}
where $\mu$ and  $\nu$ represent the initial and final state of
the system respectively.

In the next sweep of the lattice we pick $\theta_i $ at random,
attempt a Monte Carlo update of an angular variable $ \theta_i
\longrightarrow \theta_i \pm \Delta \theta$ and accept or reject
it according to the Metropolis algorithm outlined above. We find
that choosing $\Delta \theta = 0.01$ allows us to equilibrate the
chain over reasonable timescales while being small enough so that
the discretization of chain conformations does not significantly
affect the numerical results. This latter point was checked by a
comparison to runs with even smaller values of the angular
variable updates. We study systems of sizes ranging from $N=10$ to
$N=100$. We see a small but finite system-size dependence in the
sharpness of the transitions of the underlying spin variables.

Finally, because of their discrete nature the spin variables
equilibrate much faster than the angular variables. In all the
runs we checked that we had performed a sufficient number of Monte
Carlo steps in order to equilibrate the largest correlated
structures in the system that have the slowest dynamics, {\em
i.e.} the equilibration time $\tau_{\rm eq} $ was always taken to
be greater than $4 \tau_{\rm cor}$ -- the largest correlation time
in the system. Due to the wide separation of the time scales for
the equilibration of the helix--coil and angular variables, a more
efficient Monte Carlo scheme could be developed by updating the
angular variables more often than the secondary structure
variables when deep in the ordered phase of these secondary
structure variables. We did not pursue such improvements of the
efficiency of the code there.

We benchmarked our code by comparing our results to the known
equilibrium properties of the Ising model and the Kratky-Porod
model in the limit where we have artificially frozen either the
tangent vectors or the spin variables respectively. In these cases
the HCWLC model reduces to these well-studied cases.

\begin{figure}
\includegraphics[width=7cm]{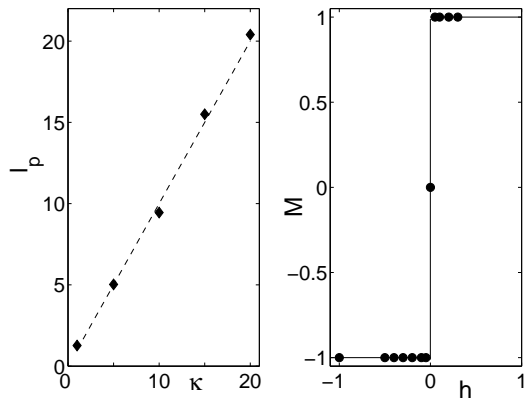}
\caption{\label{mvshlpmc} a) Variation of the persistence length
as a function of bending modulus of the WLC with $\kappa$ measured
in units of $k_{B} T$. b) The Ising limit of the model showing the
magnetization curve: $M$ vs $h$ at $ k_{\rm B} T=10^{-4}$. Both
the figures are for an $N=20$ chain. The error bars are the size
of the symbols.}
\end{figure}

As an example of these checks of the numerics we first examine the
WLC limit of the model by fixing the secondary structure variables
$s_i = +1$ for all $i$. In this limit, the HCWLC Hamiltonian is
given by Eq.~\ref{WLC-Hamiltonian} up to a trivial constant. The
correlation function for this WLC Hamiltonian can be calculated
exactly\cite{Fisher:63}. For our tangent vectors in two dimensions
we find that
\begin{equation}
\label{WLC-correlation-length} \langle \hat{t}_{i} \cdot
\hat{t}_{j} \rangle \sim \exp[- | j-i |/l_{p} ],
\end{equation}
where $l_{p}$ is the persistence length of the chain that is given
in terms of the bending modulus by
\begin{equation}
\label{WLC-persistence-length}
l_{p} = - \log[ \frac{1}{2 \pi  I_{1}(\kappa)}],
\end{equation}
This length is the distance along the chain over which tangent
vectors decorrelate. We have equilibrated the chain over $2 \times
10^4$ time steps (corresponding to about $500$ lattice sweeps)
before making our measurements. We show in panel a) of the figure
Fig.\ref{mvshlpmc} the dependence of the persistence length upon
the bending modulus. The dashed line shows the agreement to the
prediction of  Eq.~\ref{WLC-persistence-length}. In
panel b) of the same figure we examine the Ising limit of the
model in which we fix the tangent vectors to lie along the
$\hat{x}$-axis. Here we plot  the zero-temperature phase
transition of the helix-coil (Ising) variables as a function of
the solvent quality $h$ (magnetic field in the Ising language). As
predicted for the one-dimensional Ising model, the system jumps
spontaneously to the helix phase upon making $h$ an arbitrarily
small positive quantity.

\section{Results}
\label{results}
\subsection{The radius of gyration}
\label{radius-gyration}

We now turn to the full HCWLC model and discuss first the radius
of gyration\cite{Benoit:53}. To do so we note that the projection
of the polymer arc length along the average tangent vector of a
segment, {\em i.e.} the effective length of the segment depends on
the state of secondary structure. To account for this aspect of
the coarse-grained HCWLC polymer model we define a segment length
that is a function of the secondary structure variable $s_n$ via
\begin{equation}
\label{gamma-def} \gamma(s) = \left\{ \begin{array}{ll}
                        \gamma_< & \mbox{if $s = + 1$} \\
                        \gamma_> & \mbox{if $s=-1$}
                   \end{array}
            \right.  ,
\end{equation}
where, as the notation suggests, $\gamma_< < \gamma_>$. The length
of a segment increases when it loses its alpha-helical secondary
structure. We return to a discussion of reasonable numerical
estimates of these values in the conclusions.

The separation vector between the $i^{\rm th}$ and $j^{\rm
th}$ segments along the chain is given by
\begin{equation}
\label{Rij} R_{ij} = \sum^{j-1}_{n=i} \gamma (s_n) \hat{t}_{n},
\end{equation}
where $\gamma (s_n)$ is the length of the $n^{\mbox{th}}$ segment
measured along its mean chain tangent as defined above.

\begin{figure}
\includegraphics[width=7cm]{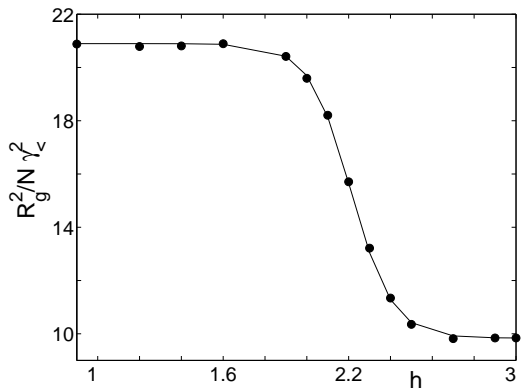}
\caption{\label{rgmc} Radius of gyration of the HCWLC as a
function of the free energy cost per segment to transform to the
random coil, nonnative state: $h$ obtained from Monte Carlo
simulations (plotted using black dots) as compared with the theory
\protect{\cite{Chakrabarti:04}}. In this curve $\kappa_{>} = 100$,
$\kappa_{<} = 1$, $N = 10$, and $\epsilon_{\rm w} = 10$. All the
energy scales are measured in $k_{B} T$ and the error bars are
smaller than the symbols in the plot.}
\end{figure}

To compute the radius of gyration in our simulations we first
evaluate the center of mass of the polypeptide. This is given by
the expression
\begin{equation}
\label{cm-defn} \vec{R}_{\rm cm} = \frac{1}{N}
\sum^{N}_{j=1}\sum^{j}_{i=1} \gamma (s_{i}) \hat{t}_{i}.
\end{equation}
The radius of gyration is then evaluated by computing the average
\begin{equation}
\label{rg-defn} R^{2}_{\rm G} =  \frac{1}{N} \langle \sum^{N}_{j=1}
\left(  \sum^{j}_{i=1} \gamma (s_{i}) \hat{t}_{i} - \vec{R}_{\rm cm}\right)^2 \rangle .
\end{equation}
We plot in Fig.~\ref{rgmc} the radius of gyration as a function of
$h$ for $\epsilon_{\rm w}=10 $, and bending moduli $\kappa_> =
100$, and $\kappa_<=1$ for an $N=10$ sized chain obtained using
Monte Carlo simulations (black dots). These data are in excellent
agreement with our analytic calculations (solid line).  The energy
scale $h$ is the free energy cost of a segment being in its
nonnative ({\em i.e.} random coil) state and so variations of $h$
mimic changing the solvent quality as is done in the
denaturation of proteins by the addition of {\em e.g.}
Guanidinium\cite{Arakawa:84}. A second effect of solvent quality
is the change in the effective excluded volume interaction between
segments of the polymer. This is not included here but, since the
chain never approaches a random coil configuration in these
simulations we expect that the absence of self-avoidance plays a
minimal role in the above results.

Similar behavior is seen in the dependence of  the squared
averaged end--to--end vector of the chain. For high chain
cooperativities and $h$ chosen so that the chain is in either the
all helix or all coil phase we find this quantity has the
following expected scaling form for a worm-like chain: $\Delta R^2
= N^2 \gamma^2 g(\frac{N \gamma }{l_{p}})$, where we have
introduced the function: $g(x) = 2 ( \exp[-x] - 1 + x )/x^2$ and
$\gamma$ is either $\gamma_>$ or $\gamma_<$ for the all helix or
all coil chains respectively. The behavior of the HCWLC is similar
to the WLC but with a persistence length that interpolates between
that of the helix and the random coil. This results also agrees
with previous analytic calculations\cite{Chakrabarti:04}.

\subsection{Mechanical Properties}
\label{mechanical-properties}

\subsubsection{Torque response}
\label{bending} To begin our numerical exploration of the
mechanical properties of the model, we consider the response of
the chain to externally applied torques in thermal equilibrium.
These torques act to constrain the tangent vectors of the ends of
the molecule while not applying tensile stress. Since the
restricted partition function of the chain with arbitrarily fixed
end tangents can be computed exactly, we expect the results of the
simulations to agree with the previous work for all values of the
model parameters. The response of the chain
to tensile stress or to the combination of tensile stress and
bending torques, however, cannot be computed analytically in
closed form; it is there that the complementary utility of the
Monte Carlo approach will be seen.

To study the torque response of the chain, we hold the first
tangent vector fixed along the $\hat{x}$-axis and constrain the
last chain tangent to make a fixed angle $\psi$ with respect to
the same axis.  We numerically evaluate the constraining torque
$\tau(\psi)$ by computing the derivative of the free energy with
respect to the angle $\psi$. In our simulations this is effected
by directly calculating the thermal average:
\begin{equation}
\label{torq-estimator}
\tau(\psi) = \langle \kappa(s_{N-1}) \sin[\psi -
\theta_{N-1}] \rangle.
\end{equation}

\begin{figure}
\includegraphics[width=7cm]{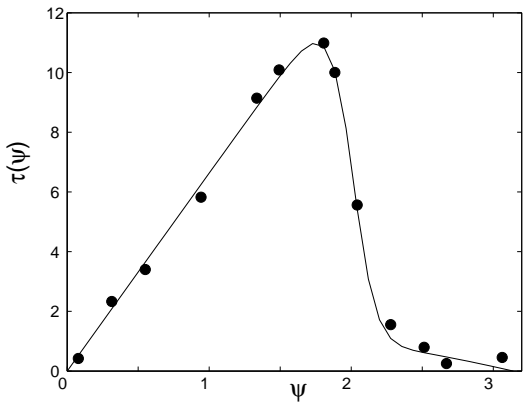}
\caption{\label{torqmc} Numerical torque vs.\ angle data (filled
points) for $\epsilon_{\rm w} \approx 5.2$, $h=8$,
$\kappa_{>}=100$, and $\kappa_{<}=1$ compared with previous exact
calculations [13,14] (solid line) for an $N=10$ chain. The torque
is measured in units of $k_{B} T$. The error bars are of the same
size as the symbols in the plot.}
\end{figure}

The results, which corroborate the analytic calculations are
plotted in figure \ref{torqmc}. At small values of the bending
angle $\psi$, there is a linear dependence of the constraining
torque on $\psi$. The alpha--helix bends like a flexible, elastic
rod. At a certain critical angle $\psi^\star$, however, the
constraining torque reaches a maximum and then drops precipitously
for angles $\psi > \psi^\star$ as shown in figure~\ref{torqmc}.
This dramatic collapse of the chain's rigidity is akin to the
buckling instability of a macroscopic tube such as a drinking
straw. The mode of the localized failure though is quite
different. In the case of current interest, the failure is caused
by the localized disruption of the secondary structure. The
breaking of the hydrogen bonds at this denatured site introduces a
weak link allowing the molecule to bend at a lower torque. At
$\psi = \psi^\star$, $M$, the fraction of the chain in the
nonnative state, abruptly jumps to $\mathcal{O}(1/N)$
demonstrating that, within the model, the buckling failure is due
to the creation of a single random coil segment along the chain
that provides a region of greatly reduced bending stiffness. The
size of the created random-coil section will remain on the order
of $N \kappa_</\kappa_>$ so for a large difference in bending
moduli between the native and nonnative states of the chain, these
``weak links'' generically occupy a small fraction of the polymer.
For instance, in the example shown in figure~\ref{torqmc} there is
only one weak link created.

The above behavior is seen in the parameter range ($\epsilon_{\rm
w}$, $h$) consistent with the chain being in an all helix state in
thermal equilibrium in the absence of applied torque. If the
equilibrium system is in a mixed helix-coil phase such a behavior
is not observed. Instead we see a monotonic increase in the
bending torque for all end angles less than $ \pi$ as expected for
an elastic rod. At $\psi = \pi$ the torque measured in thermal
equilibrium is identically zero since one is averaging a signed
quantity and the statistical weight of positive and negative
torques becomes equal. The average of the squared torque, however,
vanishes only at $\psi = 0$ modulo $2 \pi$.
Since it is reasonable to suppose that the chain bends in the
plane defined by first and last constrained chain tangents, one
expects that our two-dimensional results accurately captures the
buckling instability of the chain. We expect that the full three
dimensional calculation would generate at least qualitatively
similar results.

We now turn to the numerical study of the extensional compliance
of the chain. These Monte Carlo simulations complete the analysis
of extensional compliance of the model by accessing parameter
regimes where neither of our previous approximations are valid. In
particular we look at the extensional compliance where the mean
field approximation fails and we look at the modification of the
extensional compliance due to applied torque. The existence of
this torque/stretch coupling is due to the inherent nonlinearity
of the model.

\subsubsection{Extensional compliance}
\label{stretching}

In the presence of a tensile force $F$, the Hamiltonian of the
HCWLC may be written as
\begin{equation}
H = H_{0} - F \sum^{N}_{i=0} \gamma(s_{i}) \cos(\theta_{i}),
\label{Hamiltonian-stretching}
\end{equation}
where $H_{0}$ is the HCWLC Hamiltonian in the absence of
externally applied forces as shown in Eq.~\ref{HCWLC-hamiltonian}.
In our Monte Carlo simulations we apply a force of magnitude $F$
and then equilibrate the chain as described above. We then
calculate the quantity
\begin{equation}
\label{chain-length} L(F) = \langle \sum^{N}_{i=1} \gamma(s_i)
\hat{t}_{i} \rangle.
\end{equation}
This result gives the mean length of the chain as a function of
the externally applied force. Similar measurements are made for
the case where the initial and final chain tangents are
constrained: $\theta_1 = 0$ and $\theta_N = \psi$.  These latter
simulations allow us to explore the nonlinear dependence of the
extensional compliance upon applied torque. In such simulations we
compute
\begin{equation}
\label{chain-length-bend} L(F,\psi) = \left. \langle
\sum^{N}_{i=1} \gamma(s_i) \hat{t}_{i} \rangle\right|_{\theta_0 =
0, \theta_N = \psi}.
\end{equation}
In all the extension measurements the force is always collinear
with the initial chain tangent, $\hat{t}_1 = \hat{x}$.

\begin{figure}
\includegraphics[width=7cm]{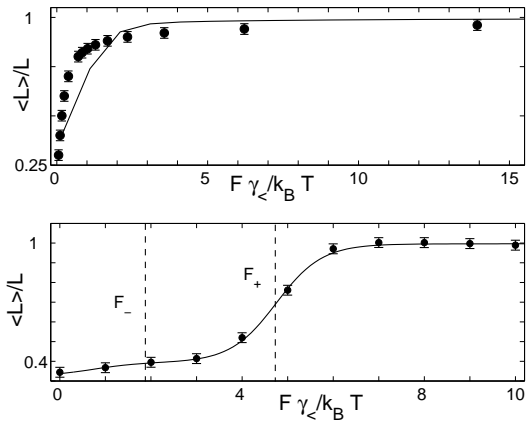}
\caption{\label{fvsextmcfinal} Force vs.\ extension Monte Carlo
data (points) shown in the low (upper panel) and high (lower
panel) cooperativity limit.  In each the mean length of the chain
normalized by the maximum chain length $N \gamma_{>}$ is plotted
as a function of the applied force normalized by the length of a
helix segment $\gamma_{>}$. For the upper panel the parameters are
$\epsilon_{w} = 0.5$, $h=1.0$, $\kappa_{>}=4.0$, $\kappa_{<}=2.0$,
and $N=20$.  The lower panel the parameters are $\epsilon_{\rm
w}=8$, and $h=1.5$, and $\kappa_{>}= 100$, $\kappa_{<}=1$, and
$N=10$. In both graphs the mean field calculation
\protect{\cite{Chakrabarti:04}} is shown as a solid line.}
\end{figure}

We plot in figure~\ref{fvsextmcfinal} the force-extension curve of
the HCWLC model with two sets of parameter values. In both force
extension measurements we do not constrain the final chain
tangents and numerically determine the function defined in
Eq.~\ref{chain-length}. In the first case (upper panel of the
figure) we have selected the HCWLC model parameters such that the
polymer is in a fluctuation--dominated regime where the mean-field
approximation is not valid. In the lower panel we have tuned those
parameters to suppress secondary structure fluctuations.

In upper panel we take $\epsilon_{\rm w}=0.5$ and $h=1.0$ to
enhance the secondary structure fluctuations. The bending moduli
are chosen to be $\kappa_> = 4$ and $\kappa_< = 2$ respectively.
Here we do not expect the mean field theory (shown as a solid line
in figure~\ref{fvsextmcfinal}) to fit the simulation data. The
mean field theory indeed fails to describe the force extension
curve adequately. In particular we note that the most significant
discrepancy between the mean-field theory and the data occurs at
low forces. In the limit of high forces, the tension in the
polymer acts as an ordering field suppressing the fluctuations
that the mean-field theory ignores.  Moreover, we note that the
intermediate plateau (``pseudoplateau'') in the graph of extension
vs.\ applied force vanishes for this parameter regime. We return
to the vanishing of the pseudoplateau below.  In fact in the limit
of low chain cooperativity, the polymer is typically found in a
mixed helix/coil phase and we obtain force extension curves
similar to those the WLC results of Marko and
Siggia\cite{Marko:95}.

In the limit of high chain cooperativity (simulation data and mean
field prediction shown in the lower panel of
figure~\ref{fvsextmcfinal}), the data agree well with our previous
mean-field calculations using the HCWLC model as expected. In
this case we have taken $\epsilon_{\rm w} = 8$ to enforce high
chain cooperativity and suppress secondary structure fluctuations.
We have taken the remaining parameters to be $h=1.5$,
$\kappa_{>}=100$ and $\kappa_{<} = 1$.

Returning to the appearance of the pseudoplateau in the upper
panel of figure~\ref{fvsextmcfinal} we recount the explanation of
its existence. For small applied forces, a molecule in the
all-helical phase having large stiffness approaches its maximum
length $N \gamma_{<}$ as $1/\sqrt{\frac{F \kappa_{>} \gamma_<}{k_B
T}}$. The WLC predicts the vanishing of the extensional compliance
as the chain reaches its maximal length; this would produce a
plateau at these intermediate forces. This plateau, however, is
not flat in the HCWLC due to the fact that an increase in the
applied force enhances the fluctuations into the longer, random
coil phase of the segments, which we refer to as the
pseudoplateau. The end of this plateau is marked by a sharp
lengthening transition at a value of force $F_+ \approx
\frac{\epsilon_{\rm w} + h}{\Delta \gamma}$ where $\Delta \gamma =
\gamma_> - \gamma_<$. Here we observe the force-induced
denaturation of helical domains.

Upon closer examination one notes that the appearance of this
intermediate-force pseudoplateau is actually a finite-size effect
in the model. The width of this pseudoplateau is controlled by two
forces, a force at onset $F_{-}$ and a maximum force $F_+ $ at
which point the abrupt denaturation transition occurs. These
forces  are marked on figure~\ref{fvsextmcfinal}. The lower force
$F_{-}$ is determined by the beginning of the first Marko-Siggia
plateau where the amplitude of transverse thermal undulations of
the all alpha-helical polymer have been significantly reduced.
Because that reduction has an algebraic dependence on applied
force, it is not possible to unambiguously select a critical force
marking the onset of the plateau. It appears reasonable, however,
to insist that plateau has been reached when $dL/dF \,
(1/\gamma_<^2) \ll 1$, {\em i.e.} the when the incremental
extension of the chain measured in monomer lengths $\delta
L/\gamma_<$ is small for a change in force $\delta F \sim
1/\gamma_<$ set by the thermal energy ($k_{\rm B} T = 1$) divided
by the same monomer size. In this case we find the force
associated with the plateau onset $F_{-} \sim N^{2/3}
\kappa_>^{-1/3} \gamma_<^{-1}$ grows with the length of the chain.
Longer chains require larger forces to sufficiently pull out the
equilibrium population of transverse undulations and thereby reach
the plateau regime.

The high force end of the plateau is determined by balancing the
free energy cost associated with transforming a segment from alpha
helix to random coil with the work done by the external force
during that transformation: $ h + \epsilon_{\rm w} \sim \Delta
\gamma F_+$. On the left hand side is the free energy cost of
making one domain wall and converting one segment to the
thermodynamically unfavored random coil state. The right hand side
of the same relation gives the work done by the external force as
one segment lengthens by  $\Delta \gamma $. The existence of the
pseudoplateau requires that $F_{-} < F_+$; the applied force must
straighten out the alpha helix at tensions small enough such that
the force-induced helix-to-coil transition does not occur. This
final inequality demands that
\begin{equation}
\label{plateau-criterion} N \le \left( \frac{h + \epsilon_{\rm
w}}{ \gamma_< \, \Delta \gamma } \right)^{3/2} \kappa_>^2 .
\end{equation}
Typical alpha helical polypeptides are quite short $N \sim {\cal
O}(10)$ and stiff $\kappa_> \sim {\cal O}(10^2)$ so that one
expects to observe this pseudoplateau behavior quite generally. We
see that for the parameters used to create the upper panel of
figure~\ref{fvsextmcfinal} the criterion for the presence of the
pseudoplateau is not met and in our Monte Carlo simulations we
indeed observe no pseudoplateau for these values. When the
criterion is met as in the case shown in the lower panel of figure
\ref{fvsextmcfinal}, the mean field example, the pseudoplateau is
evident.

We now consider the force extension measurements made while
constraining the initial and final chain tangents. In this manner
we compute the mean extension as a function of applied force as
well as the imposed total angular bend on the polymer. This
quantity is defined by Eq.~\ref{chain-length-bend}. In the regime
where the persistence length of the chain in both the helix and
the coil states is smaller than the chain itself, these
constraining torques have essentially no impact on the extensional
compliance of the chain. The effect of the angular constraints is
limited to boundary regions within a persistence length of the
constrained ends. For stiffer chains having a persistence length
longer than the chain itself, there is a significant nonlinear
coupling between these constraining torques and the extensional
compliance of the polymer at low to moderate forces.

\begin{figure}
\includegraphics[width=7cm]{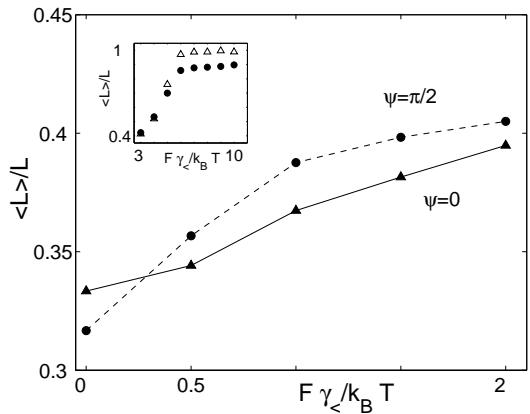}
\caption{Force vs.\ extension curves for HCWLC chain with
constrained end tangents. The force vs.\ extension  is shown for
chains where the first tangent lies along the direction of the
pulling force while the second tangent makes an angle of $\psi =
0$ (triangles) or $\psi = \pi/2$ (circles) with respect to the
initial chain tangent. The parameters of the chain are:
$\kappa_{>}=40$ $\kappa_{<}=2$, $\epsilon_{\rm w}=10$, $h=1$, and
$N=20$. Note the extensional compliance depends on the state of
the final chain tangent. The inset shows that the chain extension
in the direction of applied force saturates at high forces to
different values as a result of the constrained end tangent.
\label{forcetorquecouplefinalmc}}
\end{figure}

Such effects are shown in figure \ref{forcetorquecouplefinalmc}
where we plot two force extension curves for a HCWLC having
parameters $\kappa_{>}=40$, $\kappa_{<}=2$, $\epsilon_{\rm w}=8$
and $h=1.5$. In both cases the initial chain tangent is
constrained to lie along the $\hat{x}$-axis, the direction of the
tensile force. In one case (triangles) the final chain tangent is
collinear with the initial chain tangent, while in the second case
(circles) the final chain tangent makes an angle $\psi = \pi/2$
with the initial one. These simulations numerically evaluate the
thermal average of the chain's extension as defined in
Eq.~\ref{chain-length-bend} with $\psi = 0, \pi/2$.

At low applied force we see that, due to the nonlinear coupling of
applied torque and force, the chain has a greater extensional
compliance in the bent state. The $\psi=\pi/2$ chain reaches its
pseudoplateau more rapidly with increasing force. Once the chain
denatures under the tensile stress causing the persistence length
to become shorter than the chain itself, the coupling between the
extensional compliance and end tangent constraint disappears.  As
one approaches arbitrarily high forces the effect of the
constrained end tangent itself becomes localized at the last
segment of chain regardless of the bending moduli of the polymer
so that its effect becomes negligible.  In the limiting case of
infinite force there remains only the effect of the constrained
end itself, which decreases the length of the chain by $\gamma_>
(1- \cos \psi)$. In the strongly fluctuating secondary structure
regime, we have not been able to extract any systematic
information about the nonlinear coupling of force and torque for
low forces.

The numerical calculation presented here does not directly probe
the force/torque coupling since we have measured the extensional
compliance in an ensemble of chains with fixed end tangents. We
have directly measured the function defined by
Eq.~\ref{chain-length-bend}. Clearly, the constraint torques
applied to the end have a complicated dependence on the
extensional stress. In order to quantitatively probe the
force--torque coupling it is necessary to work in a fixed torque
ensemble rather than a fixed angle one. The results presented here
may, however, directly apply to protein mechanics where allosteric
rotations of the final tangents of an alpha helical domain can
directly modify the extensional elasticity of the helix thus
affecting both its mean length and fluctuation spectrum.

Returning to the observed breakdown of the mean field theory for
the force extension curves, we ask can one develop a simple
criterion to predict the region of HCWLC parameter space where our
analytic mean field analysis will hold?  Indeed from the energy
fluctuations observed in the Monte Carlo simulation one can
directly map the boundaries of the fluctuation-dominated regime.
It is clearly desirable to develop a criterion that can be written
directly in terms of the model parameters so that one may know
immediately where the simpler analytic theory applies.

We observe that if the polymer were straight the effect of the
force on the remaining secondary structure variables would be to
simply shift the effective free energy cost per segment to destroy
the secondary structure from $h$ to $h_{\rm eff} = h - F \Delta
\gamma$; the cost of denaturing the segment is $h$ but the net
work done by the chain upon extension under the external force $F$
is $- F \Delta \gamma$.  We recall that the validity of the mean
field theory requires only the suppression of secondary structure
fluctuations. We then expect that mean field theory to fail where
$h_{\rm eff} = h - F \Delta \gamma$ is small, which for reasonable
values of $h$ requires significant forces. Assuming that these
forces have quenched most of the contour fluctuations of the
chain, we take as an approximate criterion for the breakdown of
mean field theory the Ginzburg criterion for the one-dimension
helix-coil model with an effective field $h_{\rm eff}$. Thus mean
field theory is expected to hold when
\begin{equation}
\label{Ginzburg-criterion} \frac{< s^2_{i} > - < s_i >^2}{< s_i
>^2} \ll 1.
\end{equation}
Neglecting boundary effects to restore the translational
invariance along the chain and taking $N$ large enough so that we
may consider only the $\lambda_1$, larger of the two eigenvalues
of the transfer matrix, this condition can be written as
\begin{equation}
\label{G-criterion-eig} \frac{\lambda_1 (\partial^2
\lambda_1/\partial h^2)}{(\partial \lambda_1/\partial h)^2} \ll 1
\end{equation}
where the larger eigenvalue, computed in \cite{Chakrabarti:04} is
given by
\begin{eqnarray}
\lambda_1 &=& (1 + e^{-h_{\rm eff}}) + \nonumber \\
&  &\frac{1}{2} \sqrt{(1 -e^{- h_{\rm eff}})^2 + 4 \exp[ - 2 \epsilon_{\rm w}- h_{\rm eff}]}
\label{Larger-eig},
\end{eqnarray}
where $h_{\rm eff}$ is defined above.

\begin{figure}
\includegraphics[width=7cm]{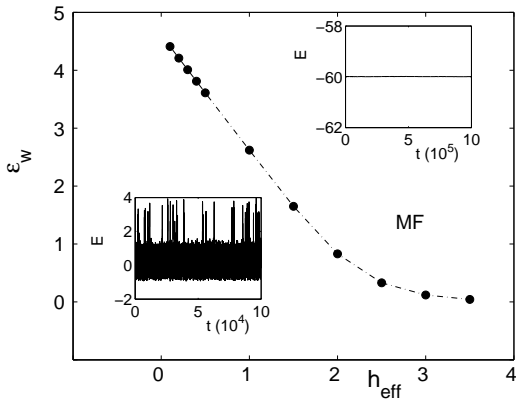}
\caption{\label{Gcriterion} A map of the $\epsilon_{\rm w}-h_{\rm
eff}$ parameter space of the helix coil degrees of freedom showing
the region of validity of the mean field approximation (MF) as
determined by the Ginzburg criterion. The insets show the energy
fluctuations over Monte Carlo steps in numerical simulations of
the two regimes.}
\end{figure}

We determined the boundary marking the limits of validity of the
mean field theory in the parameter space spanned by $\epsilon_{\rm
w}$ and $h_{\rm eff}$ is shown in Fig.\ref{Gcriterion}. To the
right of the dots mean field theory holds while in the region to
the left of this division the system is dominated by large
secondary structure fluctuations. Two traces of total energy as a
function of Monte Carlo time representative of each region are
shown as insets. In figure~\ref{Gcriterion} the filled points
represent the boundary at which the energy fluctuations reach one
part in $10^3$ of the mean. The Ginzburg criterion defined by Eq.
\ref{G-criterion-eig} give the dashed line when the left hand side
of Eq.~\ref{G-criterion-eig} is also taken to be $10^{-3}$. The
correspondence of the Ginzburg criterion and the Monte Carlo
fluctuation data demonstrates the utility of this measure of the
validity of the mean field approximation. The Ginzburg criterion
as described above correctly distinguishes these two regimes. From
that figure it is clear that mean field theory holds in the limit
of high chain cooperativity as expected. We note, however, that
since the force-induced denaturation of the chain requires $h_{\rm
eff} \longrightarrow 0$ we expect that at this transition there
will be significant fluctuation effects. Such effects have not
been fully explored.

\section{Conclusions}
\label{conclusions}

We have numerically explored the nonlinear elastic response of the
helix/coil worm-like chain under applied forces and torques. Our
Monte Carlo simulations of the extensional compliance of the
polymer agree satisfactorily with previous analytic calculations
in two limits: (i) for small forces our results are consistent
with perturbative calculations, and (ii) where mean field theory
is expected to hold as determined by a Ginzburg criterion, we find
that our mean field analysis is consistent with the Monte Carlo
data. It should also be noted that the response of the polymer to
applied torque agrees with the analytic calculations; this is to
be expected as the force-free partition function of the system can
be computed exactly in closed form.

The appropriate energy scales $h$ and $\epsilon_{\rm w}$ to
describe typical alpha-helical polypeptides under physiological
conditions can be estimated to be $h \approx 1.5$ and
$\epsilon_{\rm w} \approx 7$ \cite{Yang:95,Chakrabartty:94}. For
such parameters and polypeptides of no more than $N \sim {\cal
O}(10)$ turns it appears from our Ginzburg criterion that the
chain is typically well-described by the previous mean field
approximation. Nevertheless in order to understand the statistical
mechanics of force-induced denaturation of long polypeptides where
$\Delta \gamma \, F \sim h$ or the denaturation via changes in
solvent quality so that $ F =0$ and $h, \epsilon_{\rm w}
\longrightarrow 0^+$ one must study the fluctuation dominated
regime that is accessible via Monte Carlo. In this
fluctuation-dominated regime we find that the extensional behavior
is well described by the  Marko Siggia description of the WLC with
a persistence length that interpolates between its value in the
helix and coil states. In that regime the pseudo-plateau is
absent. Such Monte Carlo calculations can, in effect, probe the
unfolding transition state for such force-induced unfolding within
the current model. By further developing the model and studying
these transition states one may be able to assess the role of
chain cooperativity in the unfolding process. A better
understanding of this point should help to elucidate constant
velocity, single-protein unfolding
experiments.\cite{Leeds:03,Evans:99}.

Moreover the Monte Carlo calculations provide a nonperturbative
approach to investigating the nonlinear coupling between response
of the chain to combinations of applied tensile forces and bending
torques. Previously we had studied such nonlinear couplings only
in the limit of small forces. Our new results discussed above
allow us to explore the high force regime and reveal that at high
forces torque plays little role in determining the extensional
compliance of the molecule. However at low forces and for stiff
chains such that the persistence length of the helical segments is
comparable to, or larger than the length of the chain, the torque
plays an important role. Such high force couplings may play a role
in understanding protein conformational change.

Taken in combination with the previous analytical treatments,
these Monte Carlo calculations provide a nearly complete
description of the mechanical properties of the HCWLC model for
alpha-helical polypeptides. It remains only to develop a full,
three-dimensional description of the polymer that incorporates the
torsional degrees of freedom to the problem as well as to address
the role of chemical heterogeneity in order to develop a complete
and rather general equilibrium description of these biopolymers.

\section*{Acknowledgements}
BC and AJL thank Prof. J.\ Machta for stimulating conversations
regarding this work. BC also thanks Prof. N.\ Prokofev and B.\
Svitsunov for helpful discussions and for the use of computational
resources. BC acknowledges the hospitality of Indian Institute of
Science, Bangalore and AJL acknowledges the National Central
University, Taiwan where parts of this work were done.


\begin{thebibliography}{deGennes:79}

\bibitem{Kratky:49} O. Kratky and G. Porod, Rev. Trav. Chim. {\bf 68}, 1106
(1949).

\bibitem{Fisher:63} M. E. Fisher, Am. J. Phys. {\bf 32}, 343 (1963).


\bibitem{Smith:92} S. Smith, L. Finzi, and C. Bustamante, Science {\bf 258},
1122 (1992).

\bibitem{Perkins:94} T. T. Perkins, S. R. Quake, D. E. Smith, and S. Chu, Science {\bf 264}, 822 (1994).

\bibitem{Perkins:95} T. T. Perkins, D. E. Smith, R. G. Larson, and S. Chu, Science {\bf 268}, 83 (1995).

\bibitem{Cluzel:96} P. Cluzel, A. Lebrun, C. Heller, R. Lavery, J.-L. Viovy, D. Chatenay, F. Caron, Science {\bf 271}, 792 (1996).

\bibitem{Strick:96} T. Strick, J. Allemand, D. Bensimon, A. Bensimon, and V. Croquette, Science {\bf 271}, 1835 (1996).

\bibitem{Perkins:97} T. T. Perkins, D. E. Smith, and S. Chu, Science {\bf 276}, 2016 (1997).

\bibitem{Bustamante:03} C. Bustamante, Z. Bryant, and S. B. Smith, Nature {\bf 421}, 423 (2003).

\bibitem{Strick:03} T. R. Strick, M-N. Dessinges, G. Charvin, N. H. Dekker, J-F. Allemand, D. Bensimon and V. Croquette, Rep. Prog. Phys. {\bf 66}, 1 (2003).

\bibitem{Benoit:53} H. Benoit, and P. M. Doty, J. Chem. Phys. {\bf 57}, 958 (1953).

\bibitem{Marko:95} J. F. Marko and E. Siggia, Macromol. {\bf 28}, 8759 (1995).

\bibitem{Rief:90} M. Rief, H. Clausen-Schaumman, and H. E. Gaub, Nat.\ Struct.\ Biol.\ {\bf 6}, 346 (1990).

\bibitem{Coutier:04} T. Cloutier and J. Widom, Mol.\ Cell.\ {\bf 14}, 355 (2004).

\bibitem{Nelson:04} P. Wiggins, R. C. Phillips, and P. Nelson, cond-mat/0409003.

\bibitem{calmodulin} M. Zhang and T. Yuan Biochim. Biol. Cell. {\bf 76}, 313 (1998); J. S. Mills, and J. D. Johnson, J. Biol. Chem. {\bf 260}, 15100 (1985); B. F. Volkman, D. Lipson, D. E. Wemmer, and D. Kern, Science {\bf 291}, 2429 (2001).

\bibitem{Wriggers:98} W. Wriggers, E. Mehler, F. Pitici, H. Weinstein, and K. Schulten Biol.\ J.\ {\bf 74}, 1622 (1998).

\bibitem{Levine:04} A. J. Levine, cond-mat/0401624 (submitted to Phys. Rev. Lett.).

\bibitem{Chakrabarti:04} B. Chakrabarti, and A. J. Levine, cond-mat/0405382 (submitted to Phys. Rev. E).

\bibitem{Storm:03} C. Storm and P. C. Nelson, Phys. Rev. E.
{\bf 67}, 051906 (2003).

\bibitem{Chaikin:98} P. M. Chaikin, and T. C. Lubensky {\it Principles of
condensed matter physics}, (Cambridge University Press, 1998).

\bibitem{MacKintosh:95} F. C. MacKintosh, J. K\"{a}s, and P. A. Janmey Phys. Rev. Lett.\ {\bf 75}, 4425 (1995).

\bibitem{Lamura:01} A. Lamura, T. W. Burkhardt, and G. Gompper, Phys. Rev. E. {\bf 64}, 061801 (2001).

\bibitem{Poland:70} D. Poland, and H. A. Scheraga {\it Theory of helix-coil transitions in biopolymers; statistical mechanical theory of
order-disorder transitions in biological molecules} (Academic
Press, New York, 1970).

\bibitem{Birshtein:66} T. M. Birshtein and O. Ptitsyn, {\it Conformations of Macromolecules}, (Wiley, New York, 1966).

\bibitem{Bloomfield:99} V. A. Bloomfield, Am. J. Phys. {\bf 67}, 1212 (1999).

\bibitem{Tamashiro:01} M. Tamashiro, and P. Pincus, Phys. Rev. E. {\bf 63}, 021909 (2001).

\bibitem{Kamien:97} R. D. Kamien, T. C. Lubensky, P. Nelson, and C. S. O'Hern Europhys. Lett. {\bf 38} 237 (1997).

\bibitem{Ohern:98} C. S. O'Hern, R. D. Kamien, T. C. Lubensky, and P. Nelson Europhys. B {\bf 1}, 95 (1998).

\bibitem{Chakrabarti:04a} B. Chakrabarti, and A. J. Levine (work in progress).


\bibitem{Arakawa:84} T. Arakawa, S. N. Timasheff, Biochemistry {\bf 23}, 5924 (1984).

\bibitem{Yang:95} A-S. Yang and B. Honig, J. Mol. Biol., {\bf 252}, 351
(1995).

\bibitem{Chakrabartty:94} A. Chakrabartty, T. Kortemme, and R. L. Baldwin Protein Sci. {\bf 3}, 843 (1994).

\bibitem{Leeds:03} D. A. Smith, D. J. Brockwell, R. C. Zinober, A. W. Blake,
G. S. Beddard, P. D. Olmsted, S. E. Radford, Philosophical
Transactions of the Royal Society of London, Series A:
Mathematical, Physical and Engineering Sciences, {\bf 361}, 713,
(2003).

\bibitem{Evans:99} E. Evans, and K. Ritchie, Biophys. J. {\bf 72}, 1541
(1999).

\end{thebibliography}
\end{document}